\documentclass[aps,twocolumn]{revtex4}
\usepackage{graphics}
\usepackage{epsfig}
\usepackage{dcolumn}
\usepackage{amsmath,mathptm}
\usepackage[usenames]{color}

\begin{document}

\date{\today}

\title{Barrier crossing
to the small Holstein polaron regime}


\author{Peter Hamm$^*$, G. P.  Tsironis$^+$}

\address{$^*$Physikalisch-Chemisches Institut, Universit\"at Z\"urich, Winterthurerstr. 190,
CH-8057 Z\"urich, Switzerland\\ $^+$Department of Physics,
University of Crete and Institute of Electronic Structure and
Laser, FORTH, P.O. Box 2208, Heraklion 71003, Crete, Greece.}

\begin{abstract}
We investigate the dimensionality effects of the Holstein polaron
from the fully quantum regime, where the crossover between large
and small polaron solutions is known to be continuous in all
dimensions, into the limit described by the semiclassical Discrete
Nonlinear Schr\"odinger (DNLS) Equation, where the crossover is
continuous in 1D but discontinuous in higher dimensions. We use exact numerics
on one hand and a two variable parametrization of the Toyozawa ansatz on the other
in order to probe the crossover region
in all parameter regimes.  We find that a barrier appears also in
1D separating the two types of solutions, seemingly in
contradiction to the common paradigm for the DNLS according to
which the crossover is barrier-free.  We quantify the polaron behavior in the
crossover region as a function of the exciton overlap and find that the barrier remains small in
1D and tunnelling through it is not rate-limiting.
\end{abstract}

\maketitle

The Holstein model has been used widely to describe transport
properties of electrons or excitons in diverse systems, ranging
from molecular crystals~\cite{hol59} to strongly correlated
electron-phonon systems~\cite{wan98},  interface charge
localization in alkane layers~\cite{ge98},  organic transistors
~\cite{tor96,sch00}, proteins and protein-like
crystals~\cite{sco92,edl02,ham07} as well as  DNA~\cite{alex03}.
In the context of the model, the electronic degrees of freedom are
coupled to dispersionless phonons leading to the formation of
large, i.e. much larger than the lattice spacing, or small, viz.
essentially  single-site, polarons. The relative magnitudes of
three parameters of the theory, viz. electronic overlap integral,
phonon energy and exciton-phonon coupling determine the specific
polaron properties. Outstanding issues have been the nature of
polarons in different regimes, their transitions as well as the
precise onset of selftrapping. The Discrete Nonlinear
Schr{\"o}dinger (DNLS) equation, that is a semiclassical
approximation to the Holstein model, gives a continuous transition
in 1D and a discontinuous transition in higher
dimensions~\cite{kab93,kalosa98}, while most
variational calculations, in particular those
based on the Toyozawa wavefunction, show a discontinuity in all
dimensions~\cite{toyozawa61,venzl85,Zhao97,Romero98,cat99,bar02,bar07}. On the other
hand, it has been shown recently with the help of a numerically
exact solution of the full-quantum Holstein polaron that the
cross-over is continuous in 1D~\cite{bon99} as well as in higher
dimensions~\cite{ku02,ku07}. The non-existence of phase transitions in
polaron systems can also be proven on very general
grounds~\cite{gerlach87}. In the present paper, we address the
nature of the polaron transitions using exact numerics for the
fully quantum Holstein model on one hand, as well as a simple,
physically motivated two parameter variational minimization on the
other hand. The comparison of the two approaches resolves the
issue of the large to small polaron transition, produces
quantitative information on the nature of the selftrapping
transition  and finalizes pending issues regarding deficiencies of
standard approximations.

The Holstein Hamiltonian reads:
\begin{eqnarray}
H&=&H_{ex} + H_{ph} + H_{ex,ph}\nonumber\\
H_{ex}&=&  -J \sum_{j} \left({ B_j}^\dag B_{j+1} + {B_j}^\dag
B_{j-1} \right) \nonumber
\end{eqnarray}
\begin{eqnarray}
H_{ph} &=&
 \hbar\omega \sum_{j} \left({ b_j}^\dag b_j +1/2\right) \nonumber\\
H_{ex,ph}&=& -\chi \sum_{j} {B_j}^\dag B_j
 \left({ b_j}^\dag + b_j\right) \label{e1}
\end{eqnarray}
where $b_j^\dag$ ($b_j$) and $B_j^\dag$ ($B_j$) destroy (create) a
phonon and an electron (or exciton) at site $j$, respectively, and
$J$, $\hbar \omega$, $\chi$ are the excitonic overlap, phonon
energy and exciton-phonon coupling, respectively. In what follows
we use two dimensionless parameters, i.e.  the exciton coupling
$J/\hbar\omega$ and exciton-phonon coupling $\chi/\hbar\omega$
while all energies are given in units of one phonon-quantum
$\hbar\omega$. We also restrict our study to the wavenumber $k=0$
case since this will be the polaron ground state, and because it
is most relevant in optical spectroscopy ($k=0$ selection rule).

\begin{figure*}[!t]
\centerline{\epsfig{figure=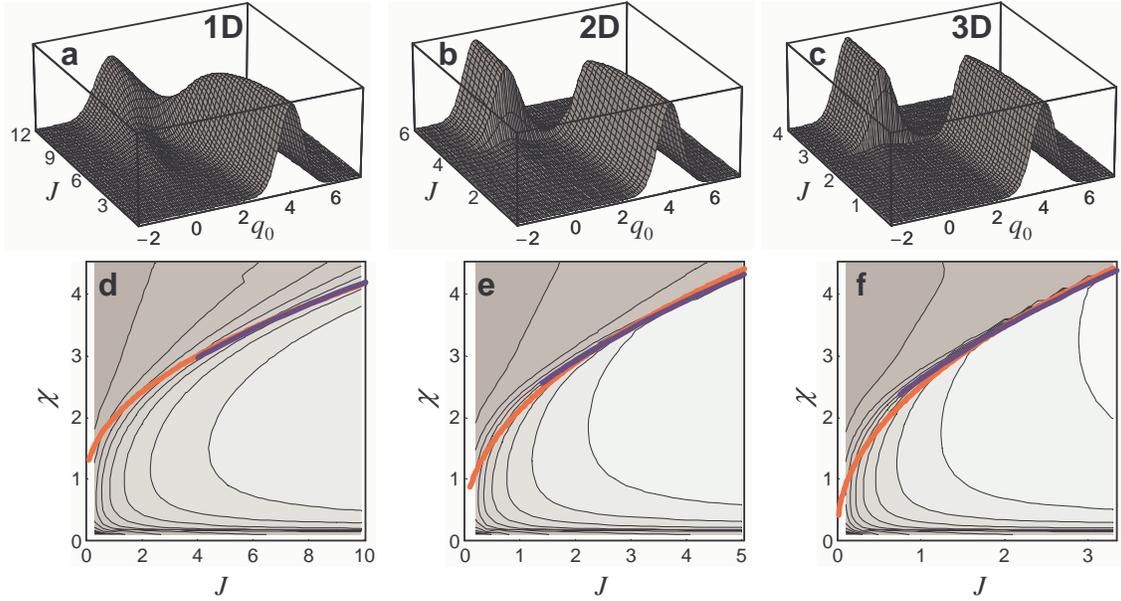,
      width=1.8\columnwidth, bbllx=40 pt,
  bblly=455 pt,bburx=550 pt,bbury=720 pt, clip=}}
\caption{ (Color online) Results from a numerically exact diagonalization of the
Holstein Hamiltonian in 1D (left), 2D (middle) and 3D (right). Top
row (a-c): Phonon density of the polaron ground state as a
function of the phonon-coordinate $q_0$. Shown is the reduced
phonon density at the site of the exciton. The exciton-phonon
coupling was set to $\chi=4.0$. Bottom row (d-f): Phase diagram,
plotting the expectation value of the phonon coordinate $\langle
q_0\rangle$ in units of $\chi$ for the polaron ground state.
Contour lines are equidistant with spacing 0.1. In red (light gray) are the
phase separation lines from Eq.~\ref{e7}, whereas in blue (dark gray) are
those obtained from equalling the two variational minima of the
Toyozawa wavefunction (see Fig.~\ref{variationalEnergy}).}
 \label{figphasediagram}
\end{figure*}

We first present numerically exact results in all three dimensions
using the approach of Trugman and coworkers~\cite{bon99,ku02}.
Figure ~\ref{figphasediagram}, top row, shows the reduced phonon
density at the site of the exciton as a function of the
phonon-coordinate $q_0\equiv (b_0^\dag+b_0)$, with phonon
coordinates at sites different from the exciton site traced out.
We note that the results are qualitatively  similar in all
dimensions: The reduced phonon density is centered at
$q_0\approx\chi$ for small exciton couplings $J$ (i.e. a small
polaron), similar to the $J=0$ case, for which an exact
(analytical) solution exists revealing with
$q_0=\chi$~\cite{sco92}. For sufficiently large exciton couplings
$J$, the phonon displacement shifts to $q_0\approx 0$, i.e.
essentially a free exciton without any phonon-displacement.  We
observe that the transition between the two regimes is smoother in
1D than it is in 2D and 3D, however, the essential point is that
free exciton and small polaron solutions \textit{coexist} in a
certain parameter range, also in 1D. The two states gradually
change their relative weights, but hardly their character, as the
exciton coupling $J$ is increased. If we interpret the reduced
phonon density as one that originates from an effective potential
$V(q_0)$, then the coexistence of two solutions hints to the
presence of a barrier separating them. This conclusion is in
disagreement with the semiclassical DNLS
case~\cite{kab93,kalosa98}, which does not reveal any barrier in
1D but rather a gradual transition.

The phase diagrams in Fig.~\ref{figphasediagram}, bottom row,
summarize these results where we plot the expectation value of the
mean displacement of the phonon coordinate $\langle q_0\rangle$ in
units of $\chi$ (which is the maximum value obtained in the $J=0$
case). For small couplings $\chi$ and $J$, the result is
qualitatively very similar in all dimensions with a continuous
transition between small and large polaron solution. Nevertheless,
it is quite evident that the transition becomes increasingly more
abrupt in 2D and 3D for large couplings (i.e. as we approach the
regime of the DNLS), whereas the cross-over stays smooth in the 1D
case. The phase separation lines can be fitted extremely well  to a
generic relationship (Fig.~\ref{figphasediagram}, bottom row, red
lines):
\begin{eqnarray}
\chi_{c}=1/\alpha+\sqrt{\alpha J}\label{e7}
\end{eqnarray}
with $\alpha=1$ in 1D, $\alpha=3.34$ in 2D and $\alpha=5.41$ in
3D, respectively. The 1D value stems from Lindenberg et al.
empirical relationship~\cite{Romero99}, whereas those for 2D and
3D coincide with the critical couplings in the DNLS limit (when
$\sqrt{\alpha J}\gg 1/\alpha$)~\cite{kalosa98}. Hence, Eq.
(\ref{e7}) does give the correct semiclassical limits expected
from the DNLS analysis.

Although exact, the results of Fig.~\ref{figphasediagram}
originate from a diagonalization of a huge matrix and provide
relatively little physical insight. We therefore employ an
approximate, yet more intuitive trial function, the Toyozawa
wavefunction~\cite{toyozawa61,venzl85, Zhao97,bar02}, and use the numerically
exact solution as an accuracy check. The Toyozawa trial function
is written as a Bloch state:
\begin{eqnarray}
|\Psi\rangle&=&\frac{1}{\sqrt N}\sum_l e^{i
kl}|\psi_l\rangle\label{e5}
\end{eqnarray}
with
\begin{eqnarray}
|\psi\rangle\equiv \sum_i a_{i} B_i^\dag|0\rangle_{ex} \prod_j
|q_{j-i}\rangle_j \label{e2}
\end{eqnarray}
where all $|\psi_l\rangle\equiv|\psi\rangle$ are identical,
$|0\rangle_{ex}$ is an exciton vacuum state and $|q\rangle_j$ a
coherent phonon state:
\begin{eqnarray}
|q\rangle_j\equiv e^{q(b_j^\dag-b_j)}|0\rangle_{ph}.
\end{eqnarray}
The Toyozawa wavefunction function can be considered a
Bloch-extension of symmetry breaking soliton
solutions~\cite{Zhao97}, where both excitons and phonons are
dressing one common trapping site $l$. In 1D, the norm of the
wavefunction $|\psi\rangle$ is~\cite{venzl85}:
\begin{eqnarray}
\langle\psi|\psi\rangle&=&\sum_{i}\left[\sum_{j}a_{j}a_{j-i}\prod_k
e^{-(q_{k}-q_{k-i})^2/2}\right]\label{e3}
\end{eqnarray}
and the expectation value of the Hamiltonian:
\begin{eqnarray}
\langle\psi|H|\psi\rangle&=&-2J\sum_{i}\left[\sum_{j}a_ja_{j-i}\prod_k
e^{-(q_k-q_{k-i+1})^2/2}\right]\label{e4}\\
&&+\sum_{i}\left[\sum_{j}a_ja_{j-i}\sum_lq_l q_{l-i}\prod_k
e^{-(q_k-q_{k-i})^2/2}\right]\nonumber\\
&&-\chi\sum_{i}\left[\sum_{j}a_ja_{j-i}(q_j+q_{j-i})\prod_k
e^{-(q_k-q_{k-i})^2/2}\right]\nonumber
\end{eqnarray}
In more than 1D (dimension $D$), the indices $i$ and $j$ are
replaced by $\{i_x,i_y(,i_z)\}$ and $\{j_x,j_y(,j_z)\}$,
respectively, in either sum/product. Furthermore, the exciton term
is repeated $D$ times with the one-site shift (i.e. the
$q_{j-i+1}$-term) in either direction (for isotropic exciton
coupling, as considered here, this can be reduced to one exciton
term with prefactor $2DJ$).

\begin{figure}[!t]
  \centerline{\epsfig{figure=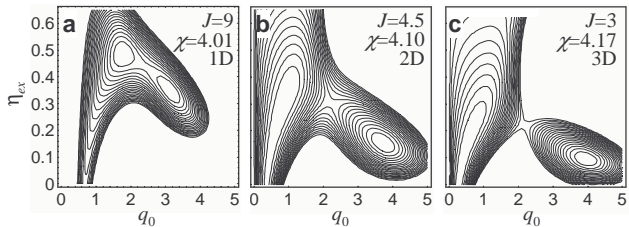,
      width=1\columnwidth, bbllx=80 pt,
  bblly=565 pt,bburx=520 pt,bbury=720 pt, clip=}}
\caption{Variational energy of the Toyozawa wavefunction as a
function of $q_0$ and $\eta_{ex}$ in 1D (left), 2D (middle) and 3D
(right). Exciton coupling is $J=9/D$ and the exciton-phonon
coupling $\chi=\chi_{c}^{var}$ has been calculated so that both
variational minima are equally deep. Contour line spacings are
$0.05$ in (a) and $0.1$ in (b) and (c).} \label{variationalEnergy}
\end{figure}

Zhao et al.~\cite{Zhao97} have minimized the energy of the
Toyozawa wavefunction by effectively varying all parameters
$\{a_i\}$ and $\{q_i\}$ (the calculation was done in momentum
space), still leaving us with a multi-parameter solution. In
contrast, we make an exponential ansatz, in analogy to
Refs.~\cite{venzl85,kalosa98,bar02}:
\begin{eqnarray}
a_i&\equiv&\eta_ {ex}^{|i|}\nonumber\\
q_i&\equiv&q_0\eta_ {ph}^{|i|} \label{e9}
\end{eqnarray}
that contains only three parameters. Minimizing Eq.~\ref{e4}
reveals $\sum_i q_i=\chi$ as one condition~\cite{venzl85,Zhao97}
which allows us to eliminate one of these parameters leading to
$\eta_{ph}=(\sqrt[D]{\chi}-\sqrt[D]{q_0})/(\sqrt[D]{\chi}+\sqrt[D]{q_0})$.
Hence, we can express the variational energy as a function of
effectively two parameters, $q_0$ and $\eta_{ex}$.

Figure~\ref{variationalEnergy} shows the variational energy of the
trial function for $J=9/D$. We observe the emergence of  two
variational minima in all dimensions with a barrier separating
them. The barrier is very shallow in 1D ($\approx0.1$) while it is
much more pronounced in 2D or 3D. This is the
barrier which is responsible for the double-peak structure of the
wavefunction in the numerically exact solution, which also exists
in 1D (Fig.~\ref{figphasediagram}a). Venzl and Fischer~\cite{venzl85} have shown
a very similar variational energy surface in 1D (as a function of
$\eta_ {ex}$ and $\eta_ {ph}$, eliminating $q_0$), however, did not address
the double-minimum due to a choice of too small value of te exciton coupling, viz.
$J=2$.

For the chosen values of the exciton-phonon coupling the two
minima are equally deep. Using the degeneracy of the variational
minima as a criterion for the critical exciton-phonon coupling
$\chi_{c}^{var}$ that induces the large-to-small polaron
transition, we find that the numerically obtained line
$\chi_{c}^{var} (J)$   follows very closely the relationship of
Eq.~\ref{e7} (compare blue and red lines in
Fig.~\ref{figphasediagram}d-f).  This near identity of
$\chi_{c}^{var}$ and Eq.~\ref{e7} starts to deviate in 1D for very
large exciton overlaps $J\gtrsim100$, indicating that Eq.~\ref{e7}
is a lower order expansion of a more complex functional
relationship.

\begin{figure}[!t]
\centerline{\epsfig{figure=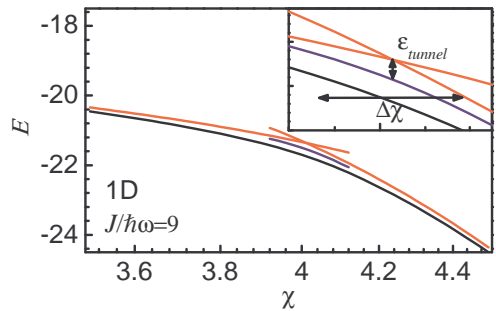,
      width=0.9\columnwidth, bbllx=40 pt,
  bblly=490 pt,bburx=560 pt,bbury=760 pt, clip=}}
\caption{(Color online) Energies of the lowest energy state for $J=9$ and the
exciton-phonon coupling $\chi$ varied through the cross-over
region in 1D. Black line: numerically exact result. Red (light gray) lines: The
two minima of the variational energy. Blue (dark gray) line: Energy corrected
by tunnelling between the two components. The inset focuses into
the crossing region and defines the terms introduced in
Eq.~\ref{e8}.} \label{fig3}
\end{figure}

Figure~\ref{fig3} shows the energy of the lowest energy
eigenstate, as deduced from the numerically exact solution (black
line), and compares it  with the two variational minima (where
they exist, red lines). Outside the crossing region, the
variational result agrees reasonably well with the exact energy --
hence the two-parameter trial function Eq.~\ref{e9} is quite good
in this regime. However, both deviate significantly at the
crossing point, indicating that the Toyozawa wavefunction
\textit{per se} does not capture the physics of the crossover
correctly~\cite{toyozawa61,Zhao97,Romero98}. In particular, the
variational minimum, as well as all other properties of the trial
wavefunction, would \textit{not} be analytical at the cross-over
point. This is simply due to the fact that the two variational
minima are separated by a barrier (Fig.~\ref{variationalEnergy}a),
and the solution of an overall minimization switches abruptly when
one of these local minima decreases below the other.

Bari\v{s}i\'{c} has extended the variational space
by considering a linear combination of two (or more) Toyozawa
wavefunctions~\cite{bar02}:
\begin{eqnarray}
|\psi\rangle\equiv c_l|\psi^{(l)}\rangle+
c_s|\psi^{(s)}\rangle\label{e6},
\end{eqnarray}
showing that this removes the discontinuity in 1D.
Fig.~\ref{variationalEnergy}a makes very clear why that is so: The
two Toyozawa wavefunctions will sit in either of two minima of the
variational energy, and tunneling coupling between them will (a)
lower the energy and thereby correct for approximately half of the
mismatch to the numerically exact result, and (b) render the
cross-over continuous (see Fig.~\ref{fig3}, blue
line)~\footnote{The expansion coefficients $c_l$ and $c_s$ in
Eq.~\ref{e6} are determined by minimization through direct matrix
diagonalization. The overlap and mixing matrix elements are
calculated in analogy to Eqs.~\ref{e3} and \ref{e4}, respectively,
with one coefficient $a_j$ or $q_k$ in either term stemming from
$|\psi^{(l)}\rangle$, and the other, $a_{j-i}$ or $q_{k-i}$, from
$|\psi^{(s)}\rangle$. Furthermore, in the 1D case, we have allowed
the positions of the two Toyozawa wavefunctions to deviate
slightly from the exact minima during the minimization, thereby
increasing the mixing matrix element on the cost of increasing the
diagonal matrix elements (energies).}

\begin{figure}[!t]
\centerline{\epsfig{figure=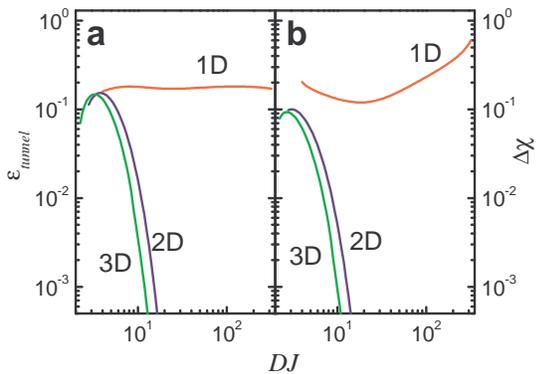,
      width=0.9\columnwidth, bbllx=30 pt,
  bblly=420 pt,bburx=560 pt,bbury=760 pt, clip=}}
\caption{ (Color online) (a) Tunnel coupling $\epsilon_{tunnel}$ 
and (b) extension
of the cross-over region $\Delta\chi$ as a function of exciton
coupling $J$ (with $\chi=\chi_{c}^{var}$) in 1D (red), 2D (blue)
and 3D (green). } \label{fig4}
\end{figure}

In order to understand the nature of the barrier separating the
small and large polaron solutions we address the question of how
the crossover domain behaves in the limit $J\rightarrow\infty$.
Mixing of the large and the small polaron states will occur
whenever the tunnel coupling is on the order of the energy
separation between them,
$\epsilon_{tunnel}\gtrsim|E^{(l)}-E^{(s)}|$. Hence, the quotient
of the tunnel coupling divided by the difference of the
derivatives of the two variational energies with respect to $\chi$
defines a measure of the extension $\Delta\chi$ of the crossover
region (Fig.~\ref{fig3}, inset, shows pictorially
$\epsilon_{tunnel}$ and $\Delta\chi$):
\begin{eqnarray}
\Delta\chi\equiv2\frac{\epsilon_{tunnel}}{d\left(E^{(l)}-E^{(s)}\right)/d\chi}\label{e8}
\end{eqnarray}
Figure~\ref{fig4}a, red line, shows that the tunnel coupling
$\epsilon_{tunnel}$ stays essentially constant in 1D  for all
$J$'s, however, the denominator in Eq.~\ref{e8} decreases with $J$
so that $\Delta\chi$ increases roughly linearly for large enough
$J$ (Fig.~\ref{fig4}b, red line). Consequently, the cross-over
remains continuous even for $J\rightarrow\infty$, leading to the
correct 1D result in the adiabatic limit. In contrast, in higher
dimensions, both the tunnel coupling $\epsilon_{tunnel}$ as well
as $\Delta\chi$ decay exponentially with $J$
(Fig.~\ref{fig4}, blue and green lines). Hence, the superposition
of two Toyozawa wavefunctions, although predicting in principle a continuous
crossover, leads (in 2D, 3D) to an increasingly more abrupt
transition for $J\rightarrow\infty$, just like the numerically
exact result does (Fig.~\ref{figphasediagram}e,f).

In conclusion, we studied the dependence of quantum polaron
regimes on lattice dimensionality employing exact numerics and a
physically motivated variational  ansatz based on the Toyozawa
wavefunction.  The exact numerical solution shows that the
large-to-small polaron crossover is continuous in all dimensions
although much more abrupt in two and three dimensions, especially
as the parameters approach the adiabatic regime
(Fig.~\ref{figphasediagram}e,f). In 1D, the transition is smooth
and remains continuous also for large $J$'s
(Fig.~\ref{figphasediagram}d).

More physical insight into the polaron features  is obtained from
an approximate trial wavefunction, viz. the Toyozawa wavefunction
of Eq.~\ref{e2}, that leads to variational results whose regime of validity
has been analyzed extensively~\cite{Zhao97,Romero98}. 
When the latter is augmented with an exponential
ansatz of Eq.~\ref{e9} containing effectively only two variational
parameters, it leads to a very intuitive and clear picture.
Despite the simplicity of this approach, it provides results that
coincide with the exact ones outside the crossover region, while,
nevertheless, predicting non-analytic properties at the transition
point in all dimensions. The appearence of this
discontinuity 
was pointed out
previously ~\cite{bar02,ku02}  but it has been found in this work
to be due to the presence of two,
almost degenerate solutions coexisting in a certain parameter
regime while separated through a barrier. Consequently, a
superposition of two Toyozawa wavefunctions, as in Eq.\ref{e6},
reveals qualitatively correct results for the cross-over region in
all dimensions and in all parameter regimes~\cite{bar02}.  The
resulting intuitive picture may be used to asses quantitatively
the properties of the cross-over region as a function of the
exciton coupling. One important outcome of this analysis is that a
barrier separating large and small polaron solutions exists also
in 1D, in sharp contrast to the predictions of semiclassical
theories. We nevertheless find that tunnelling through the barrier
remains efficient in all parameter regimes (Fig.~\ref{fig4}a, red
line), since the barrier stays of the order of $\hbar\omega$ and
hence does not hamper motion in the DNLS limit where phonons are
treated classically. The Holstein model is a fundamental
"minimal" model for strongly interacting systems and thus our
results may also have ramifications for more complex models such
as for instance the Holstein-Hubbard model.


\begin{thebibliography}{23}
\expandafter\ifx\csname natexlab\endcsname\relax\def\natexlab#1{#1}\fi
\expandafter\ifx\csname bibnamefont\endcsname\relax
  \def\bibnamefont#1{#1}\fi
\expandafter\ifx\csname bibfnamefont\endcsname\relax
  \def\bibfnamefont#1{#1}\fi
\expandafter\ifx\csname citenamefont\endcsname\relax
  \def\citenamefont#1{#1}\fi
\expandafter\ifx\csname url\endcsname\relax
  \def\url#1{\texttt{#1}}\fi
\expandafter\ifx\csname urlprefix\endcsname\relax\def\urlprefix{URL }\fi
\providecommand{\bibinfo}[2]{#2}
\providecommand{\eprint}[2][]{\url{#2}}

\bibitem[{\citenamefont{Holstein}(1959)}]{hol59}
\bibinfo{author}{\bibfnamefont{T.}~\bibnamefont{Holstein}},
  \bibinfo{journal}{Ann. Phys.} \textbf{\bibinfo{volume}{8}},
  \bibinfo{pages}{325} (\bibinfo{year}{1959}).

\bibitem[{\citenamefont{Wang et~al.}(1998)\citenamefont{Wang, Bishop, Gammel,
  and Silver}}]{wan98}
\bibinfo{author}{\bibfnamefont{W.~Z.} \bibnamefont{Wang}},
  \bibinfo{author}{\bibfnamefont{A.~R.} \bibnamefont{Bishop}},
  \bibinfo{author}{\bibfnamefont{J.~T.} \bibnamefont{Gammel}},
  \bibnamefont{and} \bibinfo{author}{\bibfnamefont{R.~N.}
  \bibnamefont{Silver}}, \bibinfo{journal}{Phys. Rev. Lett.}
  \textbf{\bibinfo{volume}{80}}, \bibinfo{pages}{3284} (\bibinfo{year}{1998}).

\bibitem[{\citenamefont{Ge et~al.}(1998)\citenamefont{Ge, Wong, {Lingle Jr.},
  McNeill, Gaffney, and Harris}}]{ge98}
\bibinfo{author}{\bibfnamefont{N.~H.} \bibnamefont{Ge}},
  \bibinfo{author}{\bibfnamefont{C.~M.} \bibnamefont{Wong}},
  \bibinfo{author}{\bibfnamefont{R.~L.} \bibnamefont{{Lingle Jr.}}},
  \bibinfo{author}{\bibfnamefont{J.~D.} \bibnamefont{McNeill}},
  \bibinfo{author}{\bibfnamefont{K.~J.} \bibnamefont{Gaffney}},
  \bibnamefont{and} \bibinfo{author}{\bibfnamefont{C.~B.}
  \bibnamefont{Harris}}, \bibinfo{journal}{Science}
  \textbf{\bibinfo{volume}{279}}, \bibinfo{pages}{202 } (\bibinfo{year}{1998}).

\bibitem[{\citenamefont{Torsi et~al.}(1996)\citenamefont{Torsi, Dodabalapur,
  Rothberg, Fung, and Katz}}]{tor96}
\bibinfo{author}{\bibfnamefont{L.}~\bibnamefont{Torsi}},
  \bibinfo{author}{\bibfnamefont{A.}~\bibnamefont{Dodabalapur}},
  \bibinfo{author}{\bibfnamefont{L.~J.} \bibnamefont{Rothberg}},
  \bibinfo{author}{\bibfnamefont{A.~W.~P.} \bibnamefont{Fung}},
  \bibnamefont{and} \bibinfo{author}{\bibfnamefont{H.~E.} \bibnamefont{Katz}},
  \bibinfo{journal}{Science} \textbf{\bibinfo{volume}{272}},
  \bibinfo{pages}{1462} (\bibinfo{year}{1996}).

\bibitem[{\citenamefont{Schoonveld et~al.}(2000)\citenamefont{Schoonveld,
  Wildeman, Fichou, Bobbert, van Wees, and Klapwijk}}]{sch00}
\bibinfo{author}{\bibfnamefont{W.~A.} \bibnamefont{Schoonveld}},
  \bibinfo{author}{\bibfnamefont{J.}~\bibnamefont{Wildeman}},
  \bibinfo{author}{\bibfnamefont{D.}~\bibnamefont{Fichou}},
  \bibinfo{author}{\bibfnamefont{P.~A.} \bibnamefont{Bobbert}},
  \bibinfo{author}{\bibfnamefont{B.~J.} \bibnamefont{van Wees}},
  \bibnamefont{and} \bibinfo{author}{\bibfnamefont{T.~M.}
  \bibnamefont{Klapwijk}}, \bibinfo{journal}{Nature}
  \textbf{\bibinfo{volume}{404}}, \bibinfo{pages}{977} (\bibinfo{year}{2000}).

\bibitem[{\citenamefont{Scott}(1992)}]{sco92}
\bibinfo{author}{\bibfnamefont{A.~C.} \bibnamefont{Scott}},
  \bibinfo{journal}{Phys. Reports} \textbf{\bibinfo{volume}{217}},
  \bibinfo{pages}{1} (\bibinfo{year}{1992}).

\bibitem[{\citenamefont{Edler et~al.}(2002)\citenamefont{Edler, Hamm, and
  Scott}}]{edl02}
\bibinfo{author}{\bibfnamefont{J.}~\bibnamefont{Edler}},
  \bibinfo{author}{\bibfnamefont{P.}~\bibnamefont{Hamm}}, \bibnamefont{and}
  \bibinfo{author}{\bibfnamefont{A.~C.} \bibnamefont{Scott}},
  \bibinfo{journal}{Phys. Rev. Lett.} \textbf{\bibinfo{volume}{88}},
  \bibinfo{pages}{067403} (\bibinfo{year}{2002}).

\bibitem[{\citenamefont{Hamm and Tsironis}(2007)}]{ham07}
\bibinfo{author}{\bibfnamefont{P.}~\bibnamefont{Hamm}} \bibnamefont{and}
  \bibinfo{author}{\bibfnamefont{G.~P.} \bibnamefont{Tsironis}},
  \bibinfo{journal}{Eur. Phys. J. Sp. Topics} \textbf{\bibinfo{volume}{147}},
  \bibinfo{pages}{303} (\bibinfo{year}{2007}).

\bibitem[{\citenamefont{Alexandre et~al.}(2003)\citenamefont{Alexandre,
  Artacho, Soler, and Chachan}}]{alex03}
\bibinfo{author}{\bibfnamefont{S.~S.} \bibnamefont{Alexandre}},
  \bibinfo{author}{\bibfnamefont{E.}~\bibnamefont{Artacho}},
  \bibinfo{author}{\bibfnamefont{J.~M.} \bibnamefont{Soler}}, \bibnamefont{and}
  \bibinfo{author}{\bibfnamefont{H.}~\bibnamefont{Chachan}},
  \bibinfo{journal}{Phys. Rev. Lett.} \textbf{\bibinfo{volume}{91}},
  \bibinfo{pages}{108105} (\bibinfo{year}{2003}).

\bibitem[{\citenamefont{Kabanov and Mashtakov}(1993)}]{kab93}
\bibinfo{author}{\bibfnamefont{V.~V.} \bibnamefont{Kabanov}} \bibnamefont{and}
  \bibinfo{author}{\bibfnamefont{O.~Y.} \bibnamefont{Mashtakov}},
  \bibinfo{journal}{Phys. Rev. B} \textbf{\bibinfo{volume}{47}},
  \bibinfo{pages}{6060} (\bibinfo{year}{1993}).

\bibitem[{\citenamefont{Kalosakas et~al.}(1998)\citenamefont{Kalosakas, Aubry,
  and Tsironis}}]{kalosa98}
\bibinfo{author}{\bibfnamefont{G.}~\bibnamefont{Kalosakas}},
  \bibinfo{author}{\bibfnamefont{S.}~\bibnamefont{Aubry}}, \bibnamefont{and}
  \bibinfo{author}{\bibfnamefont{G.~P.} \bibnamefont{Tsironis}},
  \bibinfo{journal}{Phys. Rev. B} \textbf{\bibinfo{volume}{58}},
  \bibinfo{pages}{3094} (\bibinfo{year}{1998}).

\bibitem[{\citenamefont{Toyozawa}(1961)}]{toyozawa61}
\bibinfo{author}{\bibfnamefont{Y.}~\bibnamefont{Toyozawa}},
  \bibinfo{journal}{Prog. Theor. Phys.} \textbf{\bibinfo{volume}{26}},
  \bibinfo{pages}{29} (\bibinfo{year}{1961}).

\bibitem[{\citenamefont{Venzl and Fischer}(1985)}]{venzl85}
\bibinfo{author}{\bibfnamefont{G.}~\bibnamefont{Venzl}} \bibnamefont{and}
  \bibinfo{author}{\bibfnamefont{S.}~\bibnamefont{Fischer}},
  \bibinfo{journal}{Phys. Rev. B} \textbf{\bibinfo{volume}{32}},
  \bibinfo{pages}{6437} (\bibinfo{year}{1985}).

\bibitem[{\citenamefont{Zhao et~al.}(1997)\citenamefont{Zhao, Brown, and
  Lindenberg}}]{Zhao97}
\bibinfo{author}{\bibfnamefont{Y.}~\bibnamefont{Zhao}},
  \bibinfo{author}{\bibfnamefont{D.~W.} \bibnamefont{Brown}}, \bibnamefont{and}
  \bibinfo{author}{\bibfnamefont{K.}~\bibnamefont{Lindenberg}},
  \bibinfo{journal}{J. Chem. Phys.} \textbf{\bibinfo{volume}{107}},
  \bibinfo{pages}{3159} (\bibinfo{year}{1997}).

\bibitem[{\citenamefont{Romero et~al.}(1998)\citenamefont{Romero, Brown, and
  Lindenberg}}]{Romero98}
\bibinfo{author}{\bibfnamefont{A.~H.} \bibnamefont{Romero}},
  \bibinfo{author}{\bibfnamefont{D.~W.} \bibnamefont{Brown}}, \bibnamefont{and}
  \bibinfo{author}{\bibfnamefont{K.}~\bibnamefont{Lindenberg}},
  \bibinfo{journal}{J. Chem. Phys.} \textbf{\bibinfo{volume}{109}},
  \bibinfo{pages}{6540} (\bibinfo{year}{1998}).

\bibitem[{\citenamefont{Cataudella et~al.}(1999)\citenamefont{Cataudella,
  de~Filippis, and Iadonisi}}]{cat99}
\bibinfo{author}{\bibfnamefont{V.}~\bibnamefont{Cataudella}},
  \bibinfo{author}{\bibfnamefont{G.}~\bibnamefont{de~Filippis}},
  \bibnamefont{and} \bibinfo{author}{\bibfnamefont{G.}~\bibnamefont{Iadonisi}},
  \bibinfo{journal}{Phys. Rev. B} \textbf{\bibinfo{volume}{60}},
  \bibinfo{pages}{15163} (\bibinfo{year}{1999}).

\bibitem[{\citenamefont{Bari\v{s}i\'{c}}(2002)}]{bar02}
\bibinfo{author}{\bibfnamefont{O.~S.} \bibnamefont{Bari\v{s}i\'{c}}},
  \bibinfo{journal}{Phys. Rev. B} \textbf{\bibinfo{volume}{65}},
  \bibinfo{pages}{144301} (\bibinfo{year}{2002}).

\bibitem[{\citenamefont{Bari\v{s}i\'{c}}(2007)}]{bar07}
\bibinfo{author}{\bibfnamefont{O.~S.} \bibnamefont{Bari\v{s}i\'{c}}},
  \bibinfo{journal}{Europhys. Lett.} \textbf{\bibinfo{volume}{77}},
  \bibinfo{pages}{57004} (\bibinfo{year}{2007}).

\bibitem[{\citenamefont{Bon\v{c}a et~al.}(1999)\citenamefont{Bon\v{c}a,
  Trugman, and Batisti\'{c}}}]{bon99}
\bibinfo{author}{\bibfnamefont{J.}~\bibnamefont{Bon\v{c}a}},
  \bibinfo{author}{\bibfnamefont{S.~A.} \bibnamefont{Trugman}},
  \bibnamefont{and}
  \bibinfo{author}{\bibfnamefont{I.}~\bibnamefont{Batisti\'{c}}},
  \bibinfo{journal}{Phys. Rev. B} \textbf{\bibinfo{volume}{60}},
  \bibinfo{pages}{1633} (\bibinfo{year}{1999}).

\bibitem[{\citenamefont{Ku et~al.}(2002)\citenamefont{Ku, Trugman, and
  Bon\v{c}a}}]{ku02}
\bibinfo{author}{\bibfnamefont{L.-C.} \bibnamefont{Ku}},
  \bibinfo{author}{\bibfnamefont{S.~A.} \bibnamefont{Trugman}},
  \bibnamefont{and}
  \bibinfo{author}{\bibfnamefont{J.}~\bibnamefont{Bon\v{c}a}},
  \bibinfo{journal}{Phys. Rev. B} \textbf{\bibinfo{volume}{65}},
  \bibinfo{pages}{174306} (\bibinfo{year}{2002}).

\bibitem[{\citenamefont{Ku and Trugman}(2007)}]{ku07}
\bibinfo{author}{\bibfnamefont{L.-C.} \bibnamefont{Ku}} \bibnamefont{and}
  \bibinfo{author}{\bibfnamefont{S.~A.} \bibnamefont{Trugman}},
  \bibinfo{journal}{Phys. Rev. B} \textbf{\bibinfo{volume}{75}},
  \bibinfo{pages}{014307} (\bibinfo{year}{2007}).

\bibitem[{\citenamefont{Gerlach and L\"owen}(1987)}]{gerlach87}
\bibinfo{author}{\bibfnamefont{B.}~\bibnamefont{Gerlach}} \bibnamefont{and}
  \bibinfo{author}{\bibfnamefont{H.}~\bibnamefont{L\"owen}},
  \bibinfo{journal}{Phys. Rev. B} \textbf{\bibinfo{volume}{35}},
  \bibinfo{pages}{4291} (\bibinfo{year}{1987}).

\bibitem[{\citenamefont{Romero et~al.}(1999)\citenamefont{Romero, Brown, and
  Lindenberg}}]{Romero99}
\bibinfo{author}{\bibfnamefont{A.~H.} \bibnamefont{Romero}},
  \bibinfo{author}{\bibfnamefont{D.~W.} \bibnamefont{Brown}}, \bibnamefont{and}
  \bibinfo{author}{\bibfnamefont{K.}~\bibnamefont{Lindenberg}},
  \bibinfo{journal}{Phys. Rev. B} \textbf{\bibinfo{volume}{59}},
  \bibinfo{pages}{13728} (\bibinfo{year}{1999}).

\end{thebibliography}

\end{document}